\documentclass{article}
\usepackage{graphicx} 
\usepackage{amsmath}
\usepackage{amsfonts}
\usepackage{bm}
\usepackage{ragged2e}
\usepackage{float}
\usepackage{subcaption}
\usepackage{xcolor}

\usepackage [left = 2.5 cm, right = 2.5 cm, top = 2.5 cm, bottom = 2.5 cm, bindingoffset = 0 cm]{geometry}

\emergencystretch=\maxdimen
\title{Geometric Measure of Entanglement of Rank-2 Mixed Multiqubit States and its Estimation on a Quantum Processor}
\author{N. A. Susulovska, Kh. P. Gnatenko}
\date{}

\begin{document}

\maketitle

\begin{center}
	\emph{Ivan Franko National University of Lviv,\\
	Professor Ivan Vakarchuk Department for Theoretical Physics,\\
	12 Drahomanov St., Lviv, 79005, Ukraine}
\end{center}

\section*{Abstract}
The geometric measure of entanglement of mixed states is investigated based on its relation to spin correlations. We derive an analytical expression for the entanglement of rank-2 mixed states composed of arbitrary multiqubit pure states within a two-dimensional Hilbert subspace. The dependence of the entanglement of rank-2 mixed multiqubit states on the parameters characterizing arbitrary pure states in the mixture is studied both analytically and using quantum computing. For a particular class of mixed states, namely, equal-probability mixtures of an arbitrary multiqubit state and the Schrödinger cat state, we construct quantum protocols for detecting the geometric measure of entanglement based on its relation to spin correlations. We quantify the entanglement using both the AerSimulator and the ibm\_marrakesh quantum computer. The effects of quantum-computing errors are also analyzed. The simulation results reproduce the analytical dependencies, while measurements on the quantum device demonstrate good agreement with theoretical predictions within estimated error bounds.

\section{Introduction}

Soon after the emergence of quantum mechanics, research into the nature of entanglement has become a key focus in the field. Its roots can be traced back to the formulation of the famous EPR paradox \cite{Einstein} and its subsequent resolution through the derivation \cite{Bell} and experimental testing \cite{Aspect} of Bell's inequalities. Quantum entanglement is an elusive feature inherent to compound quantum systems, where the states of constituent subsystems exhibit non-classical correlations and cannot be described separately from each other \cite{Horodecki}. Recently, in light of the rapid progress in quantum technologies, which utilize entanglement as a fundamental resource to perform a wide range of practical tasks, studies of this phenomenon have gained renewed significance. Entangled states are essential for the implementation of various quantum information schemes, including quantum teleportation \cite{Bennet}, \cite{Bouwmeester}, quantum cryptography \cite{Ekert}, \cite{Gisin}, quantum metrology and sensing \cite{Giovannetti}, \cite{Toth}. Besides, entanglement lies at the foundation of quantum computing offering unprecedented computational capabilities unattainable for classical machines \cite{Feynman, Montanaro, Shor, Brassard, Cerezo}.

Conceptualized towards the end of the last century, primarily as devices suitable for modeling complex quantum systems \cite{Feynman}, quantum computers have since come a long way in terms of both underlying theory and physical realization. Modern quantum computers operate with superpositions of quantum states and leverage entanglement to achieve an advantage in data processing over their classical counterparts. A multitude of quantum algorithms have been proposed to tackle specific classes of complex and even classically intractable computational problems, opening an avenue for quantum supremacy \cite{Montanaro, Shor, Brassard, Cerezo}. At the core of any quantum processor is a controlled multipartite system of qubits, which can, for instance, take the form of superconducting circuits, photons, spins of atoms, depending on the engineering concept behind the device. Hence, it can be utilized as a unique experimental sandbox for studying physical properties, including entanglement, and behaviour of such systems. This means that quantum computers can be employed to investigate the entanglement of various quantum states and evaluate their potential applicability within quantum algorithms under development.

Numerous studies have addressed the problem of entanglement quantification for multipartite quantum states both analytically and in the context of quantum computing \cite{Frydryszak, Vesperini_1, Vesperini_2, Vesperini_3, Kuzmak_1, Kuzmak_2, Kuzmak_3, Gnatenko_1, Gnatenko_2, Gnatenko_3, Gnatenko_4, Susulovska, De_Simone, Wang, Mooney}. To formalize this effort, a range of entanglement measures were introduced. Among them, geometry-based measures, with their straightforward intuitive interpretation, deserve special attention \cite{Frydryszak, Shimony, Chen, Cocchiarella}. The geometric measure of entanglement, as introduced in \cite{Shimony}, quantifies the degree of entanglement of a given pure state based on its distance to the closest separable state. In \cite{Cocchiarella}, the authors rely on similar considerations to derive entanglement distance for the general case of M-qudits. A significant benefit of adopting the geometric measure of entanglement is that for systems of spins 1/2 (or qubits) it was shown to be entirely determined by observables that are easily measurable on a quantum computer \cite{Frydryszak}.  Namely, the geometric measure of entanglement of a qubit with a quantum system in a pure state can be obtained on the basis of the corresponding mean spin. Importantly, for mixed states of rank 2, this measure can be related to the values of spin correlations. This finding opens a possibility for quick experimental estimation of entanglement and potential optimization of quantum schemes by appropriate selection of entangled resource states. Such an approach comes in contrast with some other cumbersome methodologies for detecting and quantifying entanglement. For instance, authors of \cite{Wang, Mooney} employ a computationally intensive state tomography procedure to show that ring graph states are fully entangled based on the negativity measure. 

It should be noted that while the entanglement of pure states is relatively well-studied, questions related to the entanglement of mixed states as well as their role in quantum computing and communications are more nuanced \cite{Vesperini_3, Boyer, Horodecki_2, Horodecki_3, Bennett_2, Siomau}. Unlike pure states, where all quantum correlations are linked to entanglement, mixed states exhibiting non-local behaviour detached from entanglement can be found. Hence, the definitions of correlation and entanglement measures are inequivalent in the case of mixed states \cite{Vesperini_3, Horodecki_2}. Studying mixed-state entanglement remains particularly important in the context of investigations related to the interaction of quantum systems with the environment and entanglement distillation \cite{Horodecki_2, Bennett_2, Kwiat}.

The relation between the geometry-based measures and mean values of observables has recently been exploited in a range of studies to explore the entanglement of different classes of multiqubit states \cite{Vesperini_1, Vesperini_2, Kuzmak_1, Kuzmak_2, Kuzmak_3, Gnatenko_1, Gnatenko_2, Gnatenko_3, Gnatenko_4, Susulovska, De_Simone}. For example, the geometric measure of entanglement was used to quantify the entanglement of graph states of different structure, namely, evolutionary graph states with Ising interaction \cite{Gnatenko_1}, graph states generated with the action of phase shift operators \cite{Gnatenko_2, Susulovska}, graph states corresponding to directed networks \cite{Gnatenko_3}. In paper \cite{Gnatenko_4}, entanglement distance was adopted to study the entanglement of quantum states represented by bipartite graphs. The authors of \cite{De_Simone} used the same measure to quantify the entanglement of quantum states corresponding to directed graphs. The geometric measure of entanglement was also considered for special classes of rank-2 mixed states. Authors of \cite{Frydryszak} studied the entanglement of two spins in a fluctuating magnetic field analytically, while papers \cite{Kuzmak_2, Kuzmak_3} focused on developing protocols for quantification of entanglement of rank-2 mixed states on a quantum computer. The methodology was validated for the mixtures of Bell states and Schrödinger cat states, the dependency of the geometric measure of entanglement on the mixing probabilities was investigated based on the results obtained on IBM's superconducting quantum backends. In this paper, we leverage the established approach and consider a more general class of rank-2 mixed states prepared by combining arbitrary multiqubit states formed on a 2-dimensional Hilbert subspace. We derive an analytical expression for the entanglement of an arbitrary qubit with other qubits of the system in such mixed states. As an example, we consider equal-probability mixtures of an arbitrary multiqubit state from the restricted subspace and the Schrödinger cat state. For these states, the geometric measure of entanglement is evaluated using AerSimulator and the ibm\_marrakesh quantum processor. Dependence of the entanglement on the parameters characterizing the arbitrary pure state in the mixture is studied.

The paper is structured as follows. In Section 2, we provide theoretical background on the geometric measure of entanglement of rank-2 mixed states and its relation to spin correlations. We also obtain a general expression for the entanglement of rank-2 mixed states corresponding to the mixtures of arbitrary pure states from the subspace under investigation. Section 3 is dedicated to the quantification of entanglement on quantum devices. We select special cases of rank-2 mixed states obtained by mixing an arbitrary state of the multiqubit system and Schrödinger cat state with equal probabilities. For such states the geometric measure of entanglement is detected on IBM's quantum simulator AerSimulator and real quantum backend ibm\_marrakesh. The dependency of this quantity on the parameters of the arbitrary pure state present in the mixture is analyzed. Conclusions of the study are presented in Section 4.

\section{Geometric Measure of Entanglement of Rank-2 Mixed States}

Consider a system of $N$ qubits in a mixed state of rank 2 described by the density matrix

\begin{equation}
    \rho = \sum_i p_i |\psi_i\rangle \langle \psi_i |,
    \label{eq:density_matrix}
\end{equation}
where $|\psi_i\rangle$ are arbitrary pure states formed on the 2-dimensional subspace of the complete $2^{N}$-dimensional Hilbert space spanned by vectors $|\mathbf{0}\rangle = |0\rangle^{\otimes N}$, $|\mathbf{1}\rangle = |1\rangle^{\otimes N}$; probabilities $p_i$ determine the fractions of pure states $|\psi_i\rangle$ present in the mixed ensemble and satisfy condition $\sum_i p_i = 1$.  Note that under such restrictions, an arbitrary superposition state of our multipartite system from the subspace of interest can be defined similarly to that of a single qubit

\begin{equation}
    |\psi_i\rangle = \cos \frac{\theta_i}{2} |\mathbf{0}\rangle +  \sin \frac{\theta_i}{2} \text{e}^{\text{i}\phi_i} |\mathbf{1}\rangle,
    \label{eq:arb_pure_state}
\end{equation}
where state parameters can vary in the appropriate ranges $\theta_i \in [0, \pi]$, $\phi_i \in [0, 2\pi)$. Given this analogy, one can introduce operators acting on the subspace \{$|\mathbf{0}\rangle, |\mathbf{1}\rangle \}$ in the same way Pauli operators act on the state space of a single qubit \{$|0\rangle, |1\rangle \}$, and satisfying all their properties. Such operators are constructed as follows

\begin{align}
    \Sigma^x_j &= \sigma^x_0 \sigma^x_1 \dots \sigma^x_j \dots \sigma^x_{N-1}, \label{eq:pauli} \\
    \Sigma^y_j &= \sigma^x_0 \sigma^x_1 \dots \sigma^y_j \dots \sigma^x_{N-1}, \nonumber\\
    \Sigma^z_j &= I_0 I_1 \dots \sigma^z_j \dots I_{N-1}, \nonumber
    \label{eq:pauli_analog}
\end{align}
where $I_j$ is a single-qubit unity operator and $\sigma_j^\alpha$, $\alpha=x,y,$ are Pauli operators acting on qubit with index $j$ \cite{Frydryszak, Kuzmak_1, Kuzmak_2}.

In general form, the entanglement of a mixed quantum state can be calculated according to the expression

\begin{equation}
    E(\rho) = \min_{\{p_i, |\psi_i\rangle\}} \sum_i p_i E(|\psi_i \rangle),
    \label{eq:ent_mixed_general}
\end{equation}
where the minimization procedure is performed over all possible decompositions of the density matrix $\rho$  with respect to pure states, and $E(|\psi_i\rangle)$ is the entanglement of the constituent pure state $|\psi_i\rangle$. Adopting the geometric measure of entanglement as an entanglement measure for pure states, one can leverage its simple relation to the mean spin value. Namely, the entanglement of a selected qubit with other qubits in pure state $|\psi_i\rangle$ is determined as

\begin{equation}
    E_j(|\psi_i\rangle) = \frac{1}{2} \left(1 - |\langle \boldsymbol{\sigma}_j \rangle_i| \right),
    \label{eq:ent_pure_mean_spin}
\end{equation}
where $|\langle \mathbf{\boldsymbol{\sigma}}_j \rangle_i| = |\langle \psi_i | \boldsymbol{\sigma}_j | \psi_i \rangle| = \left(\langle \sigma^x_j \rangle^2 + \langle \sigma^y_j \rangle^2 + \langle \sigma^z_j \rangle^2 \right)^{1/2}$ is the mean value of spin corresponding to qubit with index $j$; $\sigma_j^x$, $\sigma^y_j$, $\sigma^z_j$ are Pauli operators acting on this qubit. 

Drawing on expressions (\ref{eq:ent_mixed_general}), (\ref{eq:ent_pure_mean_spin}), it was shown in \cite{Frydryszak} that the geometric measure of entanglement of rank-2 mixed states (\ref{eq:density_matrix}) formed on the subspace  $\{|\mathbf{0}\rangle, |\mathbf{1}\rangle \}$ is related to spin correlations. The authors obtained the following analytical expression for the geometric measure of entanglement of an arbitrary qubit with the rest of the system

\begin{equation}
    E_j(\rho) = \frac{1}{2} \left(1 - \sqrt{1 - \langle \Sigma^x_j \rangle^2 -  \langle \Sigma^y_j\rangle^2} \right).
    \label{eq:ent_mixed_correlators}
\end{equation}
Here, mean values of operators $\Sigma^\alpha_j$, $\alpha = x, y$, in a mixed state can be calculated through spin correlations as follows

\begin{equation}
    \langle \Sigma^\alpha_j \rangle = \langle\sigma^x_0 \sigma^x_1 \dots \sigma^\alpha_j \dots \sigma^x_{N-1}\rangle = \sum_i p_i \langle \psi_i| \sigma^x_0 \sigma^x_1 \dots \sigma^\alpha_j \dots \sigma^x_{N-1} |\psi_i\rangle.
    \label{eq:mean_value_mixed}
\end{equation}
Note that essentially we estimate bipartite entanglement of a single qubit with the remainder of the quantum system. This quantity can be obtained for an arbitrary qubit in the system, distinguishable by index $j$, by inserting appropriate operators $\Sigma^\alpha_j$, $\alpha = x, y$, into equation (\ref{eq:ent_mixed_correlators}). However, since the action of these operators on basis states $|\mathbf{0}\rangle$, $|\mathbf{1}\rangle$ remains the same regardless of index $j$, in our case the geometric measure of entanglement takes the same value for each qubit, and the index can be dropped in equation (\ref{eq:ent_mixed_correlators}). It should also be highlighted that the outlined approach to entanglement quantification is applicable to rank-2 mixed states decomposed in any 2-dimensional basis, as long as operators acting on this basis in the way analogous to Pauli operators can be found.

Relying on relation (\ref{eq:ent_mixed_correlators}), we can easily derive an explicit expression for the geometric measure of entanglement of mixed state (\ref{eq:density_matrix}) composed of arbitrary pure states from the subspace under investigation. According to (\ref{eq:mean_value_mixed}), mean values present in this relation take the form

\begin{align}
    \langle \Sigma^x_j \rangle &= \sum_i p_i \sin\theta_i \cos \phi_i, \label{eq:mean_x_arbitrary}\\
    \langle \Sigma^y_j \rangle &= \sum_i p_i \sin\theta_i \sin\phi_i,
    \label{eq:mean_y_arbitrary}
\end{align}
where $\theta_i$, $\phi_i$ are parameters uniquely defining pure state (\ref{eq:arb_pure_state}) mixed into (\ref{eq:density_matrix}) with probability $p_i$. Finally, substituting results (\ref{eq:mean_x_arbitrary}), (\ref{eq:mean_y_arbitrary}) into (\ref{eq:ent_mixed_correlators}) we obtain

\begin{equation}
    E(\rho) = \frac{1}{2} \left(1 - \sqrt{1 - \left(\sum_i p_i \sin\theta_i \cos \phi_i \right)^2 - \left(\sum_i p_i \sin\theta_i \sin \phi_i \right)^2} \right).\label{result}
\end{equation}
This expression shows the direct dependency between the geometric measure of entanglement of an arbitrary qubit with other qubits in a mixed state and the parameters of its constituent pure states as well as corresponding mixing probabilities.

To lay a theoretical foundation for our further study of rank-2 mixed states entanglement on the basis of quantum computations, let us investigate special cases of states (\ref{eq:density_matrix}) in more detail. Namely, we consider mixed states

\begin{equation}
    \rho^{\pm} = p |\psi^{\pm}_{\text{cat}} \rangle \langle \psi^{\pm}_{\text{cat}} | + (1 - p) |\psi(\theta, \phi) \rangle \langle \psi(\theta, \phi) |
    \label{eq:density_matrix_cat}
\end{equation}
prepared by combining N-qubit Schrödinger cat states

\begin{equation}
    |\psi^{\pm}_{\text{cat}} \rangle = \frac{1}{\sqrt{2}} \left(|\mathbf{0}\rangle \pm |\mathbf{1} \rangle \right),
    \label{eq:cat_state}
\end{equation}
and an arbitrary variational state 

\begin{equation}
    |\psi(\theta, \phi)\rangle = \cos \frac{\theta}{2} |\mathbf{0}\rangle +  \sin \frac{\theta}{2} \text{e}^{\text{i}\phi} |\mathbf{1}\rangle,
    \label{eq:arb_pure_state_no_index}
\end{equation}
$\theta \in [0, \pi]$, $\phi \in [0, 2\pi)$, from 2-dimensional subspace \{$|\mathbf{0}\rangle, |\mathbf{1}\rangle \}$ with probabilities $p$ and $1-p$, respectfully. Note that Schrödinger cat states $|\psi^{\pm}_{\text{cat}}\rangle$ are obtained from (\ref{eq:arb_pure_state_no_index}) by setting state parameters $\theta = \pi/2$ and $\phi = 0, \pi$.

In this case, spin correlations (\ref{eq:mean_x_arbitrary}), (\ref{eq:mean_y_arbitrary}) are reduced to 

\begin{align}
\langle \Sigma^x_j \rangle &= \pm p + (1-p) \sin{\theta} \cos{\phi}, \label{eq:mean_x_cat}\\
\langle \Sigma^y_k \rangle &= (1-p) \sin{\theta}\sin{\phi}, \label{eq:mean_y_cat}
\end{align}
which yields the following analytical expression for the geometric measure of entanglement

\begin{equation}
    E(\rho^\pm) = \frac{1}{2}\left(1 - \sqrt{1 - p^2 \mp 2p(1-p) \sin\theta \cos\phi - (1-p)^2 \sin^2\theta}\right).
    \label{eq:ent_mixed_cat}
\end{equation}

\section{Quantifying the Entanglement of Rank-2 Mixed States on Quantum Devices}

Correlations in equation (\ref{eq:ent_mixed_correlators}) can be straightforwardly estimated on a quantum computer based on direct measurements of mean values of Pauli strings in adherence to established quantum protocols (see, for example, \cite{Kuzmak_2, Kuzmak_3}). This allows us to experimentally quantify the geometric measure of entanglement of rank-2 mixed states (\ref{eq:density_matrix}) and track its dependency on parameters of constituent pure states and corresponding mixing probabilities. For this purpose, we utilize state-of-the-art quantum devices provided by IBM through the cloud-based IBM Quantum Platform. Namely, we conduct our studies with the help of quantum simulator AerSimulator \cite{IBM_Aer}, as well as real quantum backend ibm\_marrakesh \cite{IBM_platform}. The latter belongs to the Heron R2 generation of 156-qubit superconducting quantum processors with a heavy-hexagonal connectivity pattern. Operating with basis gates \{CZ, I, RX, RZ, RZZ, SX, X\} IBM's devices of this generation ensure high-fidelity single- and two-qubit operations and exhibit outstanding performance, making them one of the top contenders in the industry.

\begin{figure}[H]
    \centerline{\includegraphics[scale=0.45]{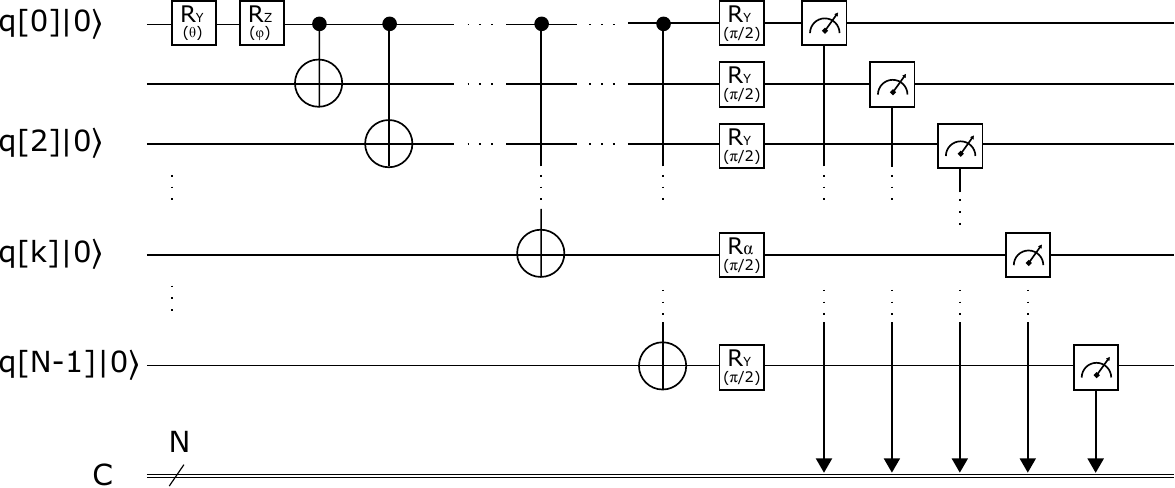}}
    \caption{A quantum protocol for determining the correlations between qubits $\langle \Sigma^\alpha_j \rangle$, $\alpha \in \{x, y\}$, in mixed state (\ref{eq:density_matrix}) on a universal gate-based quantum device. The $R\alpha$ gate rotates the state of a qubit it acts on around the respective axis $\alpha$ by angle $\pi/2$.}
    \label{fig:quantum_protocol}
\end{figure}

Expressions (\ref{eq:ent_mixed_correlators}), (\ref{eq:mean_value_mixed}) suggest that in order to estimate the degree of entanglement present in a rank-2 mixed state (\ref{eq:density_matrix}) formed on a restricted Hilbert subspace, one has to adhere to the following procedure: prepare each of the pure states comprising the mixed state of interest on a quantum device, measure spin correlations $\langle \Sigma^x_j \rangle$, $\langle \Sigma^y_j \rangle$ in each of these states and, lastly, combine the obtained results into a weighted sum with respect to mixing probabilities. At the heart of this workflow is a quantum protocol presented in Figure \ref{fig:quantum_protocol}, which can be split into blocks corresponding to state preparation and measurement of correlations.

To obtain an arbitrary pure N-qubit state (\ref{eq:arb_pure_state}) given on the subspace of interest and defined by a specific set of parameter values $\theta_i$, $\phi_i$, one of the qubits in the system q[j] is first prepared in the state from its respective state space defined by the same parameter values

\begin{equation}
    |\psi_{ij}^{init}(\theta_i, \phi_i)\rangle = \cos \frac{\theta_i}{2} |0\rangle_j + \sin \frac{\theta_i}{2} \text{e}^{\text{i}\phi_i} |1\rangle_j,
    \label{eq:init_single_qubit_arb_state}
\end{equation}
where index $j$ is used to distinguish between initial states of individual qubits when preparing entangled pure state $|\psi_i\rangle$. Afterwards, CNOT operators are consecutively applied between the selected qubit q[j] and each of the remaining qubits of the system to generate the entanglement. Conveniently, state (\ref{eq:init_single_qubit_arb_state}) can be generated on the quantum device by the action of parametrized rotation gates RY$_j$($\theta_i$), RZ$_j$($\phi_i$) on the initial state of the quantum register $|\mathbf{0}\rangle = |0\rangle^{\otimes N}$ (traditional to the majority of gate-based devices), accurate to a phase factor. Namely, the following identity holds

\begin{equation}
    |\psi_{ij}^{init}(\theta_i, \phi_i)\rangle = \text{e}^{-\text{i}\phi_i/2} RZ_j(\phi_i) RY_j(\theta_i)|\mathbf{0}\rangle,
    \label{eq:init_single_qubit_state_rotations}
\end{equation}
where single-qubit rotations are given by $RY_j(\theta_i) =\exp(-\text{i}\theta_i\sigma^y_j/2)$, $RZ_j(\phi_i) = \exp(-\text{i}\phi_i\sigma^z_j/2)$.

Further, spin correlations in the prepared pure state $|\psi_i\rangle$ can be expressed in terms of probabilities extracted from the results of measurements performed on each qubit of the system. Inherently, quantum computers only allow measurements of qubit states in the standard basis $\{|0\rangle, |1\rangle\}$ spanned by eigenvectors of Pauli operator $\sigma^z$. Since observables of interest $\Sigma^x_j$, $\Sigma^y_j$ are represented by tensor products of Pauli operators $\sigma^x$, $\sigma^y$, we leverage the following operator identities to facilitate the required measurements

\begin{align}    \label{eq:pauli_x_y_through_z}
    \sigma^x_j &= \text{e}^{-\text{i}\pi\sigma^y_j/4} \sigma^z_j \text{e}^{\text{i}\pi\sigma^y_j/4},\\
    \sigma^y_j &= \text{e}^{\text{i}\pi\sigma^x_j/4} \sigma^z_j \text{e}^{-\text{i}\pi\sigma^x_j/4}.\nonumber
\end{align}
This leads us to practical expressions for spin correlations relying solely on attainable experimental results

\begin{align}\label{eq:mean_vals_through_probs}
\langle \Sigma^x_j \rangle = \langle \psi_i|\sigma^x_0 \sigma^x_1 \dots \sigma^x_j \dots \sigma^x_{N-1}|\psi_i \rangle = \langle \Tilde{\psi}_i^{yy...y...y}|\sigma^z_0 \sigma^z_1 \dots \sigma^z_j \dots \sigma^z_{N-1}|\Tilde{\psi}_i^{yy...y...y}\rangle = p^x_+ - p^x_-,\\
\langle \Sigma^y_j \rangle = \langle \psi_i|\sigma^x_0 \sigma^x_1 \dots \sigma^y_j \dots \sigma^x_{N-1}|\psi_i \rangle = \langle \Tilde{\psi}_i^{yy...x...y}|\sigma^z_0 \sigma^z_1 \dots \sigma^z_j \dots \sigma^z_{N-1}|\Tilde{\psi}_i^{yy...x...y}\rangle = p^y_+ - p^y_-. \nonumber
\end{align}
Here

\begin{align}
    |\Tilde{\psi}_i^{yy...y...y}\rangle &= \text{e}^{\text{i}\pi(\sigma^y_0 + \sigma^y_1 + ... + \sigma^y_j + ... + \sigma^y_{N-1})/4} |\psi_i\rangle,     \label{eq:modified_states} \\
    |\Tilde{\psi}_i^{yy...x...y}\rangle &= \text{e}^{\text{i}\pi(\sigma^y_0 + \sigma^y_1 + ... - \sigma^x_j + ... + \sigma^y_{N-1})/4} |\psi_i\rangle, \nonumber
\end{align}
which suggests that before performing measurements in the standard basis, states of qubits should be rotated around $x$- or $y$-axis (depending on the corresponding Pauli operator in tensor products $\Sigma^x_j$, $\Sigma^y_j$) by angle $\pi/2$. Effectively, this means that instead of focusing on mean values of operators $\Sigma^x_j$, $\Sigma^y_j$ in the target state, we can measure mean values of product operator $\prod_j\otimes \sigma^z_j$ in states transformed according to (\ref{eq:modified_states}). Probabilities $p^{\alpha}_+$, $p^{\alpha}_-$, $\alpha=x,y$, are determined as fractions of measurement outcomes (binary strings corresponding to basis states) associated with eigenvalues $+1$, $-1$, respectfully.

This two-stage methodology is implemented for each of the pure states making up the mixed state under investigation. The number of executions of each associated quantum circuit on the quantum device is set proportional to the corresponding mixing probabilities. Finally, the correlations in the mixed state are calculated according to relation (\ref{eq:mean_value_mixed}).

In the framework of this research, we employ the approach described above to study the entanglement of a special case of rank-2 mixed states given by expression (\ref{eq:density_matrix_cat}). As previously elaborated, these states correspond to mixtures of the Schrödinger cat states (\ref{eq:cat_state}) and variational state (\ref{eq:arb_pure_state_no_index}) from the restricted subspace spanned by vectors $|\mathbf{0}\rangle$, $|\mathbf{1}\rangle$. For convenience in the context of quantum computing, we further restrict the class of mixed states under investigation by mixing two constituent pure states in equal proportions. The resulting density matrix takes the form

\begin{equation}
    \rho^{\pm} = \frac{1}{2} \left(|\psi^{\pm}_{\text{cat}} \rangle \langle \psi^{\pm}_{\text{cat}} | + |\psi(\theta, \phi) \rangle \langle \psi(\theta, \phi)| \right).
\end{equation}
Besides, we prepare our states on a system of 5 qubits.

\begin{figure}[H]
    \begin{subfigure}{0.49\textwidth}
        \centerline{\includegraphics[scale=0.55]{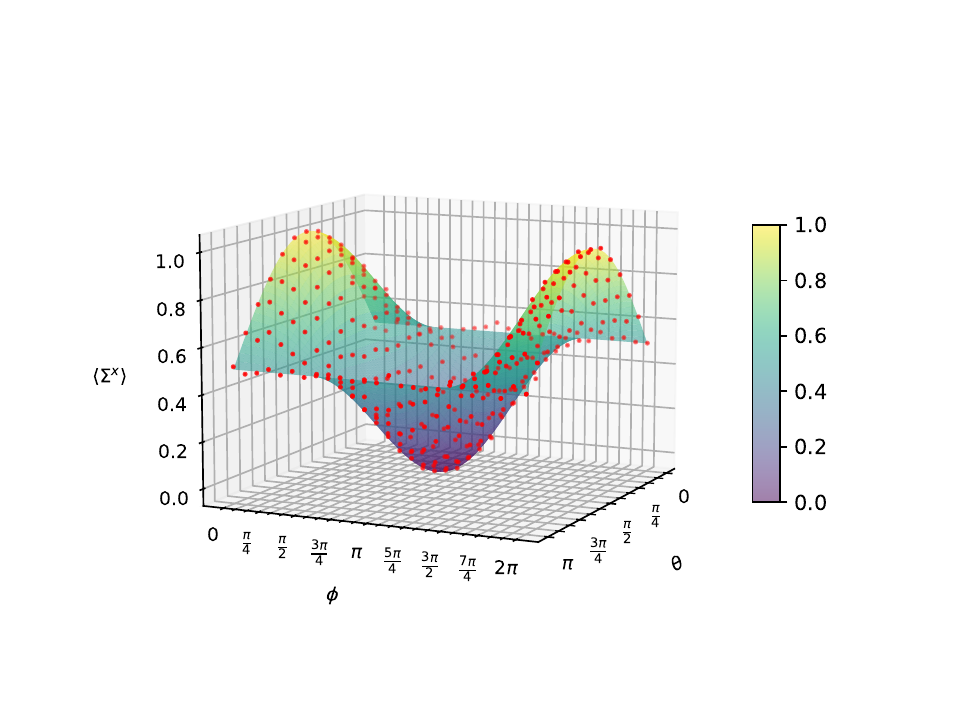}}
        \caption{}
    \end{subfigure}
    \begin{subfigure}{0.49\textwidth}
        \centerline{\includegraphics[scale=0.55]{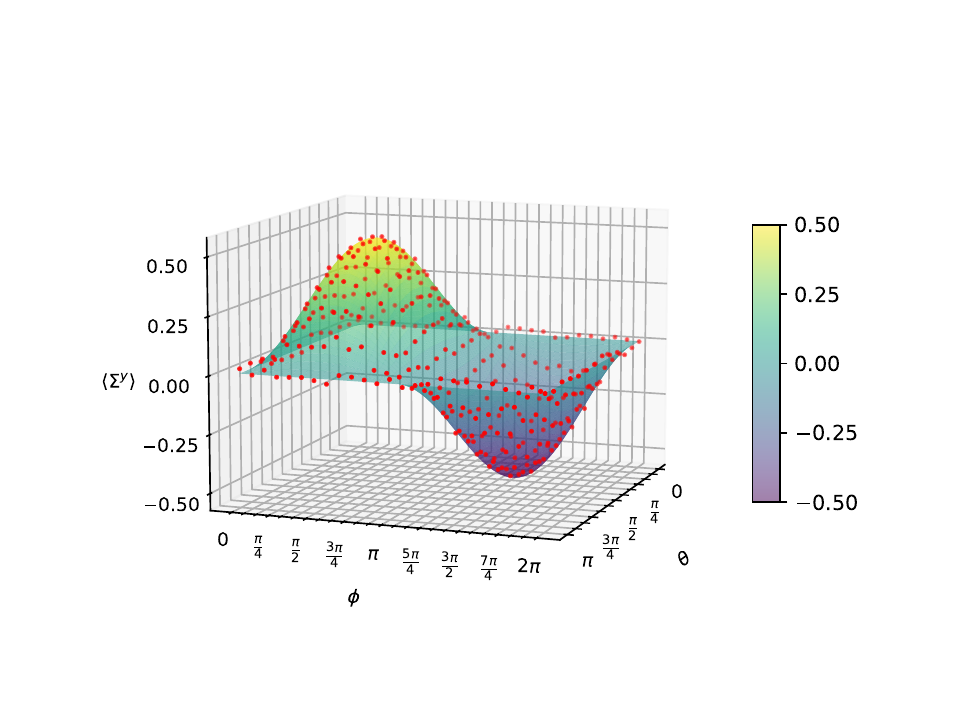}}
        \caption{}
    \end{subfigure}
    \caption{Correlators $\langle \Sigma^x_0 \rangle$ (a), $\langle \Sigma^y_0 \rangle$ (b) evaluated in mixed state (\ref{eq:density_matrix_cat}), obtained by mixing states $|\psi_{cat}^+\rangle$ and $|\psi(\theta, \phi) \rangle$ with equal probabilities. Analytical dependencies on the parameters of the constituent variational pure state (\ref{eq:arb_pure_state_no_index}) are represented with continuous surfaces. Results obtained from the ideal quantum simulation on IBM's AerSimulator for discrete parameter combinations are marked with dots.}
    \label{fig:mean_vals_sim_cat_+}
\end{figure}

\begin{figure}[H]
    \centerline{\includegraphics[scale=0.55]{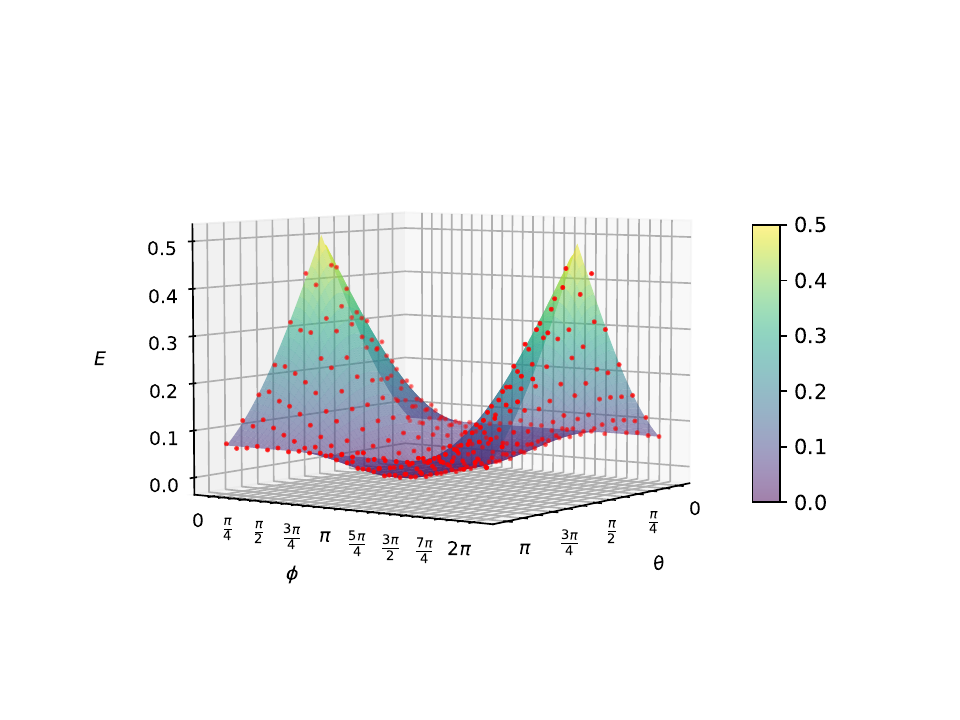}}
    \caption{Geometric measure of entanglement of qubit q[0] with the remaining system in mixed state (\ref{eq:density_matrix_cat}), obtained by mixing states $|\psi_{cat}^+\rangle$ and $|\psi(\theta, \phi) \rangle$ with equal probabilities. Analytical dependency on the parameters of the constituent variational pure state (\ref{eq:arb_pure_state_no_index}) is represented with a continuous surface. Results of quantification based on spin correlations obtained from an ideal simulation on IBM's AerSimulator for discrete parameter combinations are marked with dots.}
    \label{fig:entanglement_sim_cat_+}
\end{figure}

\begin{figure}[H]
    \centering
    \begin{subfigure}{0.49\textwidth}
        \centerline{\includegraphics[scale=0.55]{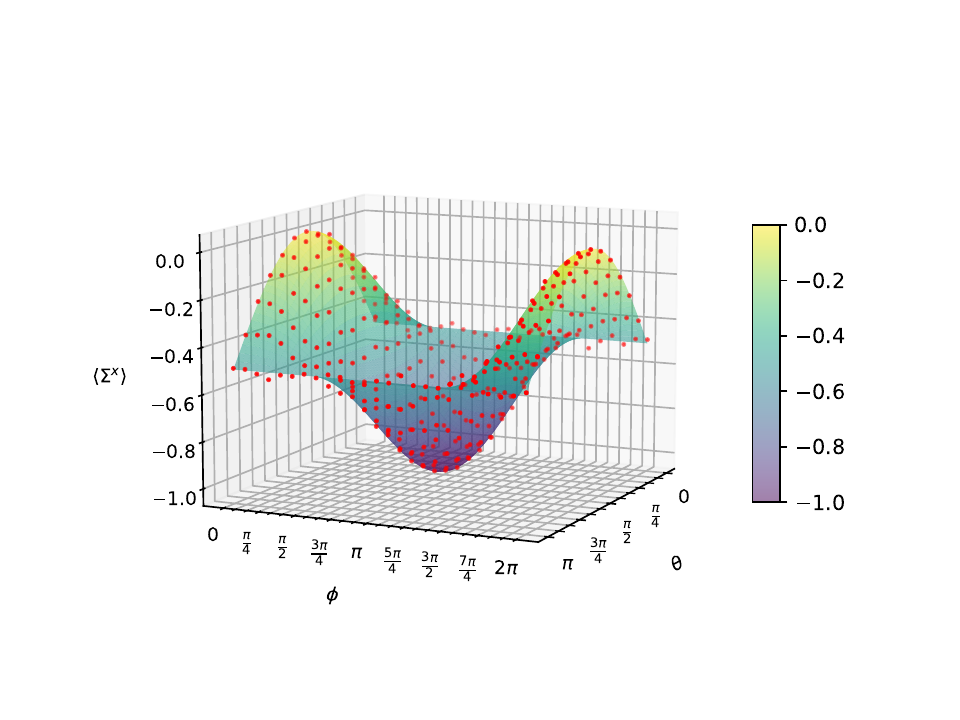}}
        \caption{}
    \end{subfigure}
    \begin{subfigure}{0.49\textwidth}
        \centerline{\includegraphics[scale=0.55]{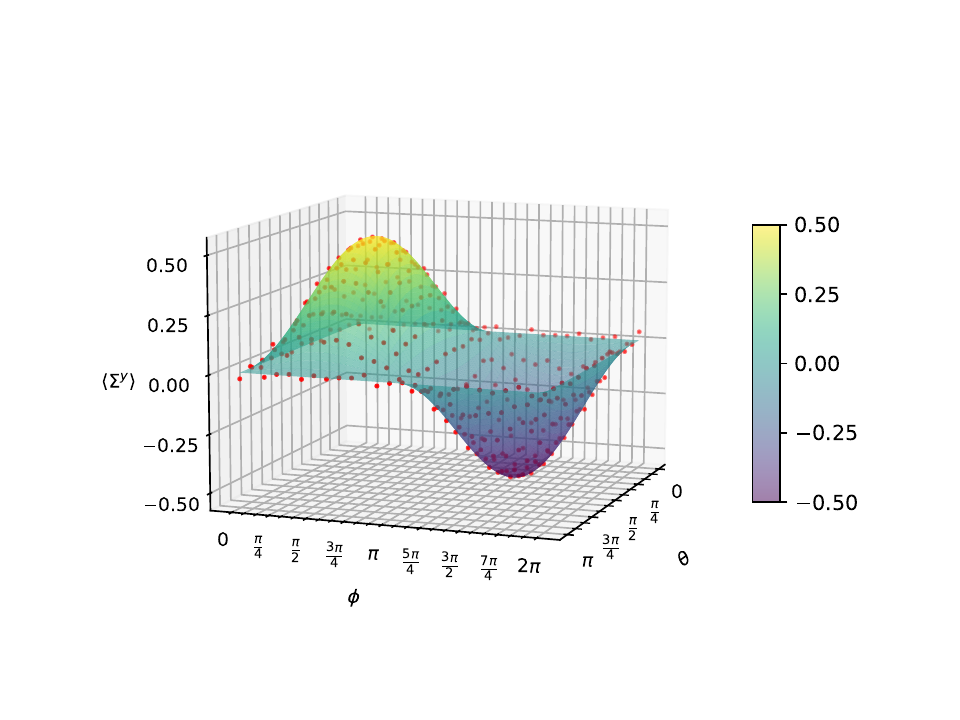}}
        \caption{}
    \end{subfigure}
    \caption{Correlators $\langle \Sigma^x_0 \rangle$ (a), $\langle \Sigma^y_0 \rangle$ (b) evaluated in mixed state (\ref{eq:density_matrix_cat}), obtained by mixing states $|\psi_{cat}^-\rangle$ and $|\psi(\theta, \phi) \rangle$ with equal probabilities. Analytical dependencies on the parameters of the constituent variational pure state (\ref{eq:arb_pure_state_no_index}) are represented with continuous surfaces. Results obtained from the ideal quantum simulation on IBM's AerSimulator for discrete parameter combinations are marked with dots.}
    \label{fig:mean_vals_sim_cat_minus}
\end{figure}

\begin{figure}[H]
    \centerline{\includegraphics[scale=0.55]{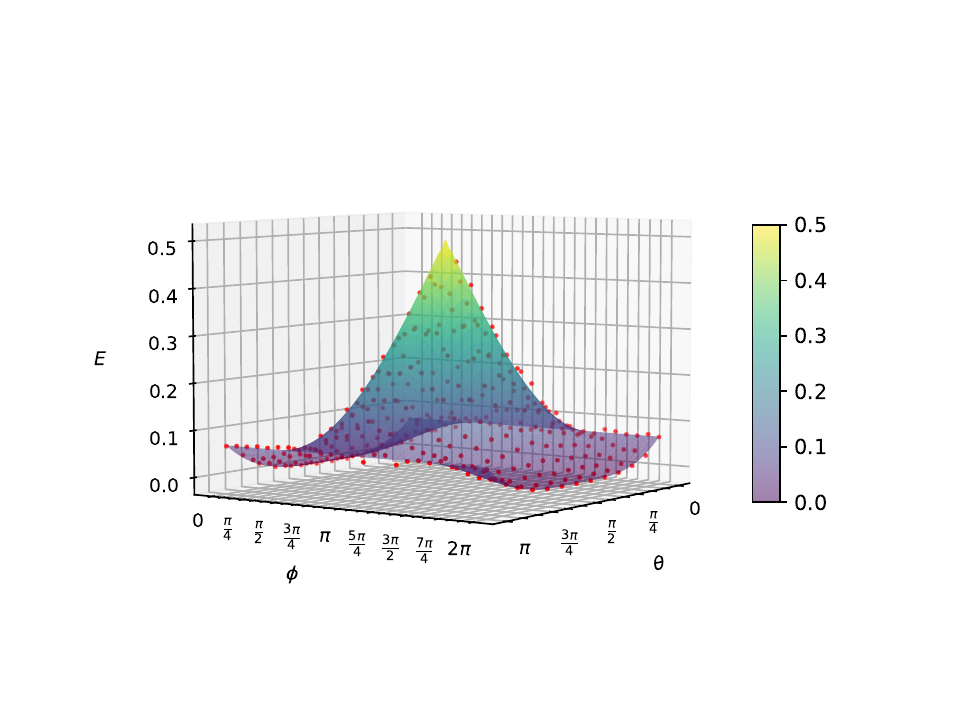}}
    \caption{Geometric measure of entanglement of qubit q[0] with the remaining system in mixed state (\ref{eq:density_matrix_cat}), obtained by mixing states $|\psi_{cat}^-\rangle$ and $|\psi(\theta, \phi) \rangle$ with equal probabilities. Analytical dependency on the parameters of the constituent variational pure state (\ref{eq:arb_pure_state_no_index}) is represented with a continuous surface. Results of quantification based on spin correlations obtained from an ideal simulation on IBM's AerSimulator for discrete parameter combinations are marked with dots.}
    \label{fig:entanglement_sim_cat_-}
\end{figure}

To begin with, the geometric measure of entanglement of a selected qubit q[0] with other qubits of the system is determined with the help of AerSimulator in the ideal setting with no simulated noise. Iterating over a discrete grid in the parameter space, where $\theta$ runs in the range $[0, \pi]$ and $\phi$ in the range $[0, 2\pi)$ with step $\pi/12$, we analyze the dependency of the spin correlations and entanglement on the parameters of the variational state present in the mixture. As expected, results of the ideal quantum simulation are well aligned with the theoretical ones. Corresponding surface plots presented in Figures \ref{fig:mean_vals_sim_cat_+}, \ref{fig:entanglement_sim_cat_+}, \ref{fig:mean_vals_sim_cat_minus}, \ref{fig:entanglement_sim_cat_-} suggest that the geometric measure of entanglement reaches its maximum magnitude of $1/2$ exclusively in case of pure states, or in other words, for the set of parameters such that the variational state coincides with the other state in the mixture, which is also consistent with theoretical knowledge. Namely, for state $\rho^+$ this situation occurs at the points $\theta=\pi/2$, $\phi=0, 2\pi$, where variational state $|\psi(\theta, \phi) \rangle$ is reduced to the Schrödinger cat state $|\psi^{+}_{\text{cat}}\rangle$, and for state $\rho^-$ at the point $\theta=\pi/2$, $\phi=\pi$, where the variational state turns into state $|\psi^{-}_{\text{cat}}\rangle$. Note that for any arbitrary mixed state, the degree of entanglement is lower than $1/2$. Entanglement vanishes, indicating a separable state, when both correlations $\langle \Sigma^x_0 \rangle$, $\langle \Sigma^y_0 \rangle$ are equal to zero.

\vspace{-1cm}

\vspace{-0.8cm}

\begin{figure}[H]
    \centering
    \begin{subfigure}{0.49\textwidth}
        \centerline{\includegraphics[scale=0.35]{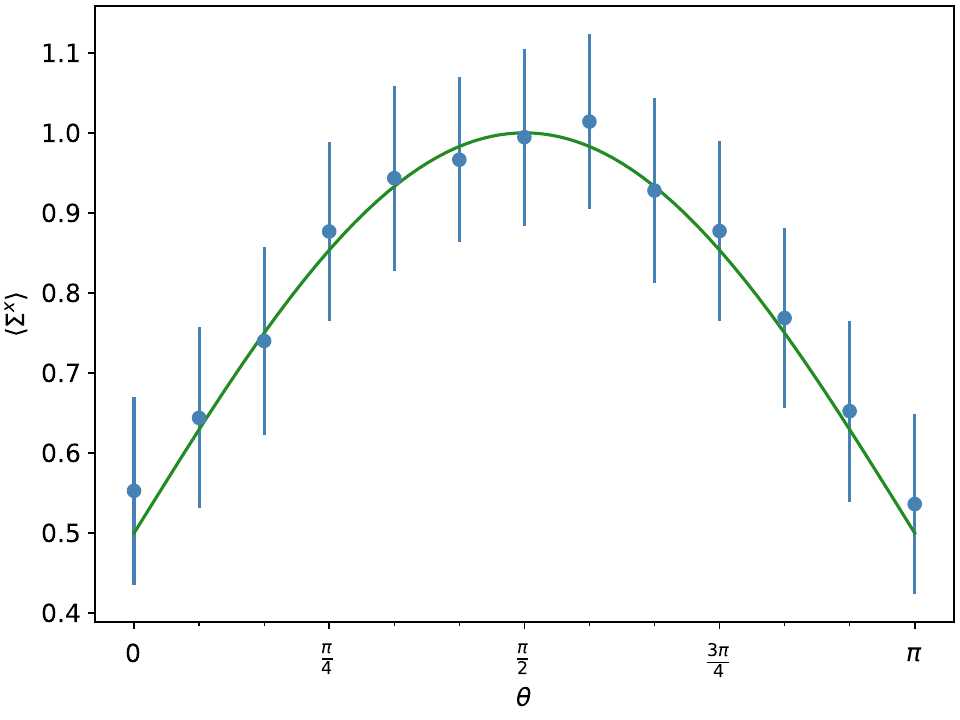}}
        \caption{}
    \end{subfigure}
    \begin{subfigure}{0.49\textwidth}
        \centerline{\includegraphics[scale=0.35]{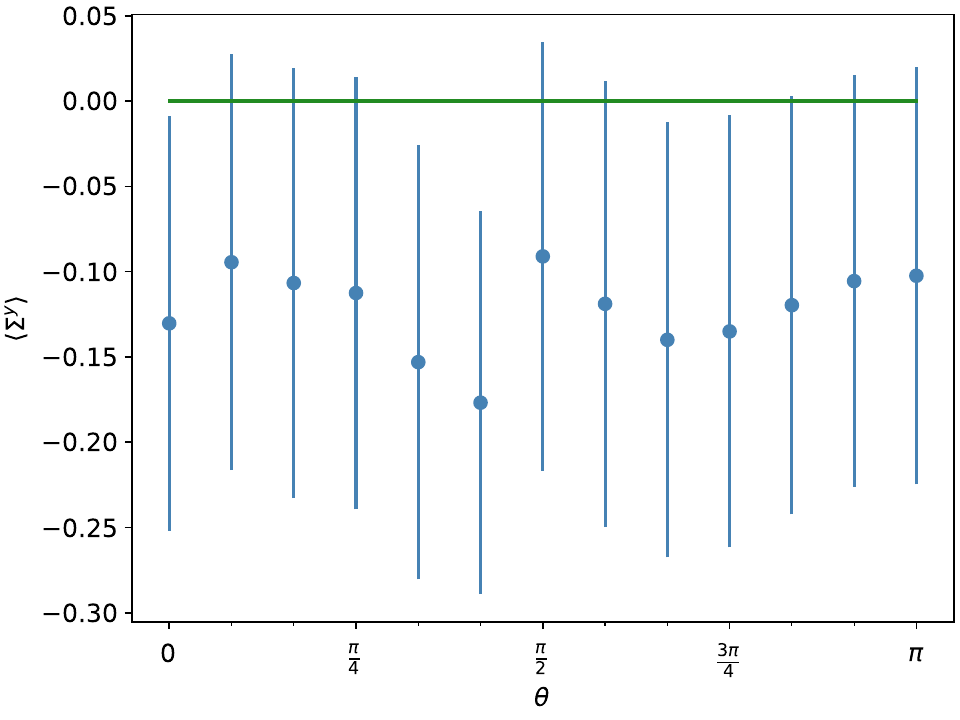}}
        \caption{}
    \end{subfigure}
    \caption{Correlations $\langle \Sigma^x_0 \rangle$ (a), $\langle \Sigma^y_0 \rangle$ (b) evaluated in mixed state (\ref{eq:density_matrix_cat}), obtained by mixing states $|\psi_{cat}^+\rangle$ and $|\psi(\theta, 0) \rangle$ with equal probabilities. Analytical dependencies on the parameter of the constituent pure state (\ref{eq:arb_pure_state_no_index}) are represented with solid lines. Results of quantification performed on IBM's quantum backend ibm\_marrakesh for discrete parameter values are marked with dots, where error bars provide cumulative uncertainty of quantum computations including contributions from statistical, gate, and readout errors.}
    \label{fig:mean_vals_qc}
\end{figure}

\begin{figure}[H]
    \centerline{\includegraphics[scale=0.35]{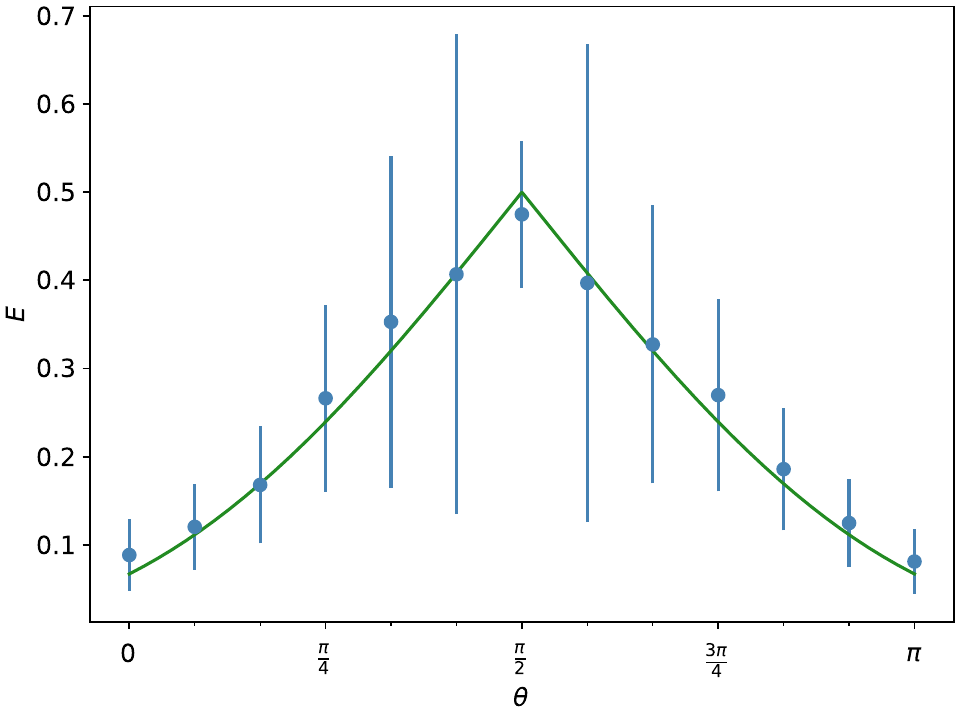}}
    \caption{Geometric measure of entanglement of qubit q[0] with the remaining system in mixed state (\ref{eq:density_matrix_cat}), obtained by mixing states $|\psi_{cat}^+\rangle$ and $|\psi(\theta, 0) \rangle$ with equal probabilities. Analytical dependencies on the parameter of the constituent pure state (\ref{eq:arb_pure_state_no_index}) are represented with a solid line. Results of quantification based on the measurements of spin correlations performed on IBM's quantum backend ibm\_marrakesh for discrete parameter values are marked with dots, where error bars provide cumulative uncertainty of quantum computations including contributions from statistical, gate, and readout errors.}
    \label{fig:entanglement_qc}
\end{figure}

Lastly, we detect the geometric measure of entanglement of rank-2 mixed states (\ref{eq:density_matrix_cat}) on IBM's quantum computer ibm\_marrakesh. In this investigation, we set relative phase $\phi$ equal to zero and observe how the spin correlations and, consequently, entanglement changes with state parameter $\theta$. Both exact and experimental dependencies are exhibited in Figures \ref{fig:mean_vals_qc}, \ref{fig:entanglement_qc}. The cumulative error of quantum computations represented on the plots by error bars is calculated as a sum of standard error, gate error and readout error. Standard error is related to the finite number of measurements performed to estimate the expectation values and describes the statistical uncertainty. It is inversely proportional to the square root of the number of shots (quantum circuit evaluations) $N_{shots}$. Since we combine pure states $|\psi(\theta, 0) \rangle$, $|\psi^{+}_{\text{cat}}\rangle$ with equal probabilities to prepare mixed state (\ref{eq:density_matrix_cat}), each of the quantum circuits used to measure spin correlations in these states is executed the same number of times, namely we assume $N_{shots} = 1024$. Gate error arises due to imperfect implementation of unitary operations on the quantum device and includes contributions from single- and two-qubit gates. Its upper limit reaches the sum of errors of each gate comprising the quantum circuit. Note that median single- (SX) and two-qubit (CZ) gate errors for ibm\_marrakesh were evaluated at $3.276 \times 10^{-4}$ and $2.782 \times 10^{-3}$, respectfully, at the time of the experiment. Readout error is caused by the bias in the measurement of quantum states, where the classical outcome is reported incorrectly. The maximum total readout error of the quantum circuit is determined as the sum of readout errors of all the qubits upon which the measurements are performed. To calculate its magnitude, we used the median readout error reported at $1.1 \times 10^{-2}$ for the utilized processor and multiplied it by the number of measurements in our quantum scheme. Note that readout error provides the most significant source of uncertainty in the results of quantum computations. Moreover, to detect the spin correlations required to calculate the entanglement of a multiqubit mixed state, all the qubits have to be measured. This means that the error grows rapidly with the increasing number of qubits in the system under consideration. However, Figure \ref{fig:entanglement_qc} shows that in the case of 5 qubits, experimental results are in good agreement with theoretical predictions, and deviations between them fall within the estimated error bounds. This serves as a benchmark of the performance of current NISQ-era quantum devices, and validates the near-term feasibility of the quantum computing approach when applied to the studies of properties of multiqubit states.

\section{Conclusions}
In this study we address the problem of entanglement quantification in the case of multiqubit mixed states, which remains highly relevant in the field of quantum information. Namely, we examine rank-2 mixed states restricted to the 2-dimensional Hilbert subspace of the complete state space of a multiqubit system spanned by vectors $|\mathbf{0}\rangle$, $|\mathbf{1}\rangle$. To determine the degree of entanglement of quantum states belonging to this class, we rely on the geometric measure of entanglement. This choice of the entanglement measure is primarily motivated by its direct relation to spin correlations. Such connection between the entanglement and observables easily measurable on quantum devices, allows us to study this property not only theoretically but also based on the empirical results of quantum computations.

For an arbitrary mixture of pure states from the restricted state space under investigation an explicit analytical expression for the geometric measure of entanglement of a qubit with the remaining system is obtained (\ref{result}). This expression establishes a relationship between the entanglement of a rank-2 mixed state and parameters characterizing its constituent pure states as well as corresponding mixing probabilities. As an example, mixtures of the Schrödinger cat state and an arbitrary state from the subspace $\{|\mathbf{0}\rangle, |\mathbf{1}\rangle\}$ are considered. An established methodology is employed to construct a quantum protocol for measuring the spin correlations required to quantify the entanglement (see Figure \ref{fig:quantum_protocol}).
The proposed protocol is applied to five-qubit mixed states formed by equal-probability mixtures of the Schrödinger cat states and variational states. Ideal simulations performed using IBM's AerSimulator reproduce the analytical dependence of the spin correlations and the geometric measure of entanglement on the parameters of the quantum states, Figures \ref{fig:mean_vals_sim_cat_+}-\ref{fig:entanglement_sim_cat_-}. We find that the entanglement reaches its maximum value of $1/2$ when the two states forming the mixture coincide, corresponding to a pure state. For any mixed state, the degree of entanglement takes a value below $1/2$. 

Finally, the entanglement of the mixtures of equal proportions of the Schrödinger cat state and a variational state is quantified on the state-of-the-art real quantum processor ibm\_marrakesh provided by IBM through the cloud-based platform. Five-qubit mixed quantum states are studied, and the relative phase of the variational state is set to zero.  The results of quantum computations are in good agreement with the theoretical predictions (see Figures \ref{fig:mean_vals_qc}-\ref{fig:entanglement_qc}). The deviations from the analytical results remain within the estimated quantum-computing errors. Overall, the outcomes demonstrate the feasibility of using direct measurements of spin correlations to quantify the geometric measure of entanglement of rank-2 mixed states on  quantum devices. 

The obtained analytical results for the entanglement of multiqubit mixed states and the proposed protocol can be used to investigate entanglement in mixed states on NISQ-quantum devices.

\section*{Acknowledgments}
This work was supported by Project 2025.07/0326 (No. 0126U002943) from National Research Foundation of Ukraine.

\end{document}